\documentclass[12pt]{article}
\usepackage[english]{babel}
\usepackage[latin1]{inputenc}
\usepackage{amsfonts,amssymb,amsmath, epsfig}
\usepackage{color,graphicx,graphics}
\usepackage{amsmath,amstext,amssymb,amsfonts, amscd}
\usepackage{hyperref}

\def\U{\Upsilon}

\def\Box{\leavevmode\vbox{\hrule
     \hbox{\vrule\kern4pt\vbox{\kern4pt}%
           \vrule}\hrule}}

\newcounter{appendix}
\def\appendix{\advance\c@appendix by 1
   \def\thesection{\Alph{section}}
   \ifnum\c@appendix=1 \setcounter{section}{-1} \fi
   \@startsection {section}{1}{\z@}{-3.5ex plus -1ex minus
   -.2ex}{2.3ex plus .2ex}{\Large\bf}}

\def\paragraph#1{{\bf #1\ }}

\newtheorem{proof}{Proof}[section]

\title{Evolution of wealth in a nonconservative economy driven by local Nash equilibria}
\author{Pierre Degond $^{(1)}$, Jian-Guo Liu$^{(2)}$, Christian Ringhofer$^{(3)}$, }
\date{}
\begin{document}

\maketitle


\begin{center}
1- Department of Mathematics,
Imperial College London, \\
London SW7 2AZ, United Kingdom. \\
email: pdegond@imperial.ac.uk
\end{center}

\begin{center}
2- Department of Physics and Department of Mathematics\\
Duke University,
Durham, NC 27708, USA\\
email: jliu@phy.duke.edu
\end{center}

\begin{center}
3- School of Mathematics and Statistical Sciences, \\
Arizona State University, Tempe AZ 85287, USA\\
email: ringhofer@asu.edu
\end{center}

\vspace{0.5 cm}
\begin{abstract}
We develop a model for the evolution of wealth in a non-conservative economic environment, extending a theory developed in \cite{Degond_etal_jsp}.
The model considers a system of rational agents interacting in a game theoretical framework.
This evolution drives the dynamic of the agents in both wealth and economic configuration variables.
The cost function is chosen to represent a risk averse strategy of each agent. That is, the agent is more likely to interact with the market, the more predictable the market, and therefore the smaller its individual  risk.
This yields a kinetic equation for an effective single particle agent density with a Nash equilibrium serving as the local thermodynamic equilibrium.
We consider a regime of scale separation where the large scale dynamics is given by a hydrodynamic closure with this local equilibrium.
A class of generalized collision invariants (GCIs) is developed to overcome the difficulty of the non-conservative property in the hydrodynamic closure derivation of the large scale dynamics for the evolution of wealth distribution.
The result is a system of gas dynamics-type equations for the density and average wealth of the agents on large scales.
We recover the inverse Gamma distribution, which has been previously considered in the literature, as a local equilibrium for particular choices of the cost function.\end{abstract}

\medskip
\noindent
{\bf Acknowledgements:} This work has been supported by
KI-Net NSF RNMS grant No. 1107291 and  DMS No. 11-07444 (KI-net).

\medskip
\noindent
{\bf Key words:} Multi-agent market models for frequent trading, volatility, price strategies, mean field games, best response strategies, inverse Gamma distribution, Pareto tail, scale separations, Fokker-Planck equation,  Gibbs measure, general collision invariants.

\medskip
\noindent
{\bf AMS Subject classification: } 91A10, 91A13, 91A40, 82C40, 82C21.
\vskip 0.4cm

\section{Introduction}
\label {intro}

\subsection {Framework}
\label {ssintro0}

\noindent
A theory on the evolution of wealth distribution driven
by local Nash equilibria
in a conservative economy was developed by the authors in \cite{Degond_etal_jsp}
in the framework set up by \cite{Degond_etal_preprint13_2},
which is closely related to Mean-Field Games
\cite{Cardaliaguet_NotesCollegeFrance12, Lasry_Lions_JapanJMath07}.
By conservative, we meant that the total wealth is preserved in the time evolution.
This assumption enabled us to derive
a large scale dynamics for the evolution of the wealth distribution by using a hydrodynamic closure with a Nash equilibrium serving
as the local thermodynamic equilibrium. This resulted in a system of
gas dynamics-type equations for the density and average wealth
of the agents on large scales. The goal of this paper is to extend this theory
to some more realistic models in non-conservative economies,
where global wealth is gained or lost at a certain rate
due to either productivity or inflation.
To overcome
the difficulty of the non-conservative property in the hydrodynamic closure,
we adapt and develop a concept of
Generalized Collision Invariant (GCI) developed by Degond and Motsch in \cite {Motsch}
for  flocking dynamics.

We consider an economy modeled as a closed ensemble of agents.
The state of each agent is described by two variables.
The variable $x$, describes its location in the economic configuration space ${\cal X}$
\cite{During_Toscani_PhysicaA07}.
 In addition, the state is described by
the wealth $y \ge 0 $ of the agent. The dynamic of these attributes is given
by some motion mechanism in the economic configuration
variable $x$ and by the exchange of wealth (trading) in the wealth variable $y$.

The subject of understanding the wealth distribution has a long history
since Pareto in 1896 \cite{Pareto_1896}. Amoroso in 1925 \cite{Amoroso_1925}
developed a dynamic equilibrium
theory and re-wrote the Pareto distribution in terms of inverse Gamma distribution.
The wealth distribution results from the combination of
two important mechanisms: the first one is the geometric Brownian
motion of finance which has first been proposed by Bachelier
in 1900 \cite{Bachelier_AnnScientENS1900} and
the second one is the trading model, one the earlier ones being
that of Edgeworth, dating back to 1881 \cite{Edgeworth_1881}.
These pioneering works have been followed by numerous authors
and have given rise to the field of econophysics.
Recent references on this problem can be found e.g.
in the books \cite{Chakrabarti_etal_2006, Naldi_etal_2010, Takayasu_2004, Takayasu_2005}
and e.g. in the references \cite{Maldarella_Pareschi_PhysA12, Silver_etal_JEconTh02,
Pareschi_Toscani_2014,
Toscani_etal_JSP13, Yakovenko_Rosser_RevModPhys09}. 
The large-scale dynamics of spatially heterogeneous social models is
currently the subject of an intense research (see e.g. \cite{Bouchaud_2014}, where the
authors investigate a spatially heterogeneous version 
of Deffuant-Weisbuch opinion model of interacting agents 
that exhibits a transition between a socially cohesive phase 
and a socially disconnected phase). 

The basic equation considered in this paper is of the form
\begin{equation}
\partial_t f (x,y,t)+\partial_x ( f \, V(x,y) ) =
 - \partial_y (f \, {\cal F}_{f} )
 + d \partial _y (\partial _y (y^{2} f))  \equiv Q(f),
\label{intro1}
\end{equation}
where $f(x,y,t)$ is the density of agents in economic configuration space $x$ having wealth $y$ at time $t$.
The second term at the right hand side of (\ref {intro1}) models the
uncertainty and has the form of a diffusion operator corresponding
to the geometric Brownian motion of economy and finance, with
variance $2dy^2$ quadratic in $y$. The justification of this operator
can be found in \cite{Oksendal_Springer}.

Here ${\cal F}_{f}$ describes the control, action or strategy.
In \cite{Degond_etal_jsp}, the authors take the action
as  the negative gradient of the cost function
$ \Phi_f$, i.e., ${\cal F}_{f}=- \partial_y \Phi_f$.
A quadratic
cost function with coefficients depending functionally
on the density $f$ was used to describe  trading behavior between agents.
We write  this cost function in general form
as
\begin{equation} \label {Phi}
\Phi _f(y)=\frac 1 2 a_fy^2+b_fy+c_f \ ,
\end{equation} with coefficients $a_f,b_f$ and $c_f$ functionally dependent
on the density $f$.

In the framework of  a non-atomic non-cooperative anonymous game with a continuum of players \cite{Aumann_Econometrica64, MasColell_JMathEcon84, Schmeidler_JStatPhys73, Shapiro_Shapley_MathOperRes78}, also known as a Mean-Field Game \cite{Cardaliaguet_NotesCollegeFrance12, Lasry_Lions_JapanJMath07}, players interact with each other to minimize their own cost function.
In this paper we consider a more realistic model, where each player
interacts with the ensemble of players, i.e. the market.
For each  player, the equilibrium reached under this interaction corresponds to the wealth difference between him/her and the market average being at one of the minima of this cost function.

We note that this model only considers the exchange of money and does not keep track of the goods and services traded. Therefore, this game does not mean that each players wishes to share some of its wealth with the trading partner. Rather, the utility of the exchange is to maximize the economic action resulting in the optimal exchange of goods and services.
Within this framework, the dynamic of agents following these strategies can be viewed as given by the following game: each agent follows what is known as the best-reply strategy, that is, it tries to minimize the cost function with respect to its wealth variable, assuming that the other agents do not change theirs.

This gives for the control action $\mathcal{F}_f$ in (\ref {intro1})
$
\mathcal{F}_f(y)=-\partial _y\Phi _f=-a_fy-b_f
$,
and for the operator $Q$ in (\ref {intro1}), including effects
of uncertainty, given by the geometric Brownian motion,
$$
Q(f)=\partial _y\big( d\partial _y(y^{2}f)+(a_fy+b_f)f\big)
$$
We consider a closed system, where the number of agents in the
market is conserved. So, equation (\ref {intro1}) is supplemented
by the boundary condition
$d\partial _y(y^{2}f)+(a_fy+b_f)f|_{y=0}=0$.

In \cite{Degond_etal_jsp, Bouchaud_Mezard_PhypsicaA00},
 a model resulting from  pairwise interactions, proportional to the quadratic distance between the wealth of the two agents is derived.
The goal of the present paper is to extend this framework to general potentials, particularly to remove
the conservation constraint for the the total wealth $\int_0^\infty y f(y,t) \, dy$.
In the following, we refer to this scenario as a "non-conservative economy".
In addition, we consider an alternative (and, in some sense, more realistic) model, where players do not interact with each other in the form of binary interactions, but with the whole ensemble of players (the market). That is, we do not consider the mean field limit of a binary interaction model, but start from an inherent mean field model.

Naturally, one takes  moments of the wealth distribution function $f$
with respect to the wealth variable $y$.
 We define the density of agents $\rho(x,t)$ and the density of
 higher order moments of the wealth variable $\rho \Upsilon _k(x,t)$, by:
\begin{equation}
\rho (x,t)=\int f(x,y,t)\ dy, \qquad
\rho \Upsilon _k(x,t)=\int y^kf(x,y,t)\ dy \ ,
k=1,2, \ldots
\label{eq:def_moments}
\end{equation}
So, $\rho (x,t)$ is the density of agents in the economic
configuration space,
$\rho \Upsilon _1(x,t)$ is the density of the mean wealth,
$\rho (\Upsilon _2-\Upsilon _1^2)$ is the density of the variance of the wealth,
and so on.
We will restrict the dependence of $a_{f}, b_{f}, c_f$ in the cost functional $\Phi _f$
to a dependence on the above defined mean densities $\Upsilon _1,\Upsilon _2, \ldots$.

\subsection {Conservative vs. non-conservative economies}
\label {ssintro1}
Computing the first three moments of the operator  $Q$ in
(\ref {intro1}) gives, using integration by parts
$$
\int \left( \begin {matrix} 1\cr y\cr y^2\end {matrix} \right) Q(f)(x,y,t)\ dy
=
 \left( \begin {matrix} 0\cr  -a_f\Upsilon _1-b_f\cr 2(d-a_f)\Upsilon _2-2b_f\Upsilon _1\end {matrix} \right) \rho (x,t)
$$
Consequently, we obtain a hierarchy for the moments of
the density function $f(x,y,t)$ with respect to the wealth variable
$y$. The first three term of the hierarchies are of the form
\begin{equation} \label {Hi}
\partial _t \left( \begin {matrix} \rho \cr \rho \Upsilon _1\cr \rho \Upsilon _2\cr \ldots
\end {matrix} \right) +
\partial _x\int V(x,y)f(x,y,t) \left( \begin {matrix} 1\cr y \cr y^2\cr \ldots\end {matrix} \right) \ dy=
 \left( \begin {matrix} 0\cr  -a_f\Upsilon _1-b_f\cr 2(d-a_f)\Upsilon _2-2b_f\Upsilon _1\cr\ldots\end {matrix} \right) \rho (x,t)
 \ .
\end{equation} The system (\ref {Hi}) is of course not closed, since the flux terms
on the left hand side of (\ref {Hi}) are in general unknown
for an arbitrary density function $f$.
The closure of  the hierarchy (\ref {Hi}) at a certain level
has to be performed by some asymptotic analysis and scaling arguments,
which are the subject of this paper.
We are faced with a conservative economy if the dependence
of the coefficients in the quadratic cost functional $\Phi _f$
on the density $f$ are such that $a_f\Upsilon _1+b_f=0$ holds
for any density $f$. In this case, the total wealth $\rho \Upsilon _1$
is preserved, when integrated over the configuration variable $x$.
So, we consider a conservative economy, for
$a_f\Upsilon _1+b_f=0$.
In this case, we would have , considering equn. (\ref {intro1}),
$\frac{d}{dt} \iint  y f(x,y,t) \, dx dy=0$, and the total wealth in the economy
would be conserved in time.

The case of a conservative economy ($a_f\Upsilon _1+b_f=0,\ \forall f$, ),
i.e. the cost functional $\Phi _f$ in (\ref {Phi})
being a parabola, centered
around $\Upsilon _1$, has been considered in
\cite{During_Toscani_PhysicaA07} and, in a game theoretical framework,
in \cite{Degond_etal_jsp}.
In this paper, we consider a non - conservative economy ($a_f\Upsilon _1+b_f\neq 0$, except in equilibrium)
where wealth is generated or lost due to productivity of the agents
or inflation.

\subsection{ Frequent trading}\label {ssft}
In this paper we will consider an asymptotic regime, where
the dynamics is dominated by the trading interaction of the
agents, i.e. where the operator $Q$ is the dominant term
in equation  (\ref{intro1}).
In the case of a conservative economy (preserving wealth with
$a_f\Upsilon _1+b_f=0,\ \forall f$),
this leads to a closed macroscopic system for the variables
$\rho $ and $\Upsilon _1$. This system has been treated in
\cite {Degond_etal_jsp} and \cite {During_Toscani_PhysicaA07}.
The more general form of the collision operator, with a general potential $\Phi _f$ in (\ref {Phi}),
still preserves the density of agents, so $1$ is a collision invariant.
(For simplicity we disregard the birth and death of the agents.)
However, the total wealth in the system is no longer necessarily
conserved if $a_f\U _1+b_f\neq 0$ holds,
although wealth is conserved in each individual transactions.
This is indeed the main driving force behind the economy and results in non - conservative economy.
The non - conservative case  considerably complicates
the derivation of a macroscopic evolution equation for the
density $\rho (x,t)$, since it is not possible to use a
local conservation
law for the mean wealth density $\rho \Upsilon _1$ in the
frequent trading limit, as done in
\cite {Degond_etal_jsp} and \cite {During_Toscani_PhysicaA07}.
We address this problem by using the concept of a general collision
invariant (GCI), as introduced in \cite {Motsch}.
This yields a macroscopic balance law (which is not conservative)
for the mean wealth density $\rho (x,t)\Upsilon _1(x,t)$ in the limit
of frequent trading.

The local equilibrium wealth distribution is also a Nash equilibrium for the non-conservative economy.
It is in general computed by solving an infinite dimensional
 fixed point problem.
However, the fixed point solution cannot be given explicitly
for general coefficients $a_f,b_f$ and $c_f$, in contrast to the previous literature where they could be expressed in terms of an inverse Gamma distribution \cite{Degond_etal_jsp}. Rather, they are found
by solving a linear partial differential equation together with a finite dimensional fixed point equation. If multiple solutions to this fixed point equation exist, corresponding to multiple stable equilibria, this indicates that phase transitions in the wealth distribution are possible. However, we leave the question of the existence and enumeration of the solutions to the fixed point equation to future work.

In Section 4 we make a particular modeling choice
for the coefficients $a_f$ and $b_f$ in the cost functional $\Phi $.
This choice corresponds to
each player interacting with the market ("trading") with a frequency
which is inverse proportional to the uncertainty of the market,
i.e. to the variation coefficient of the probability distribution
$f$ in (\ref {intro1}).
We refer to this assumption as the "risk averse" scenario,
which means that traders are more likely to trade, the better they
can predict the development of the market.
In addition, each player
tries to achieve an acceptable risk level
(given by a constant $\kappa $
which has to be matched to actual market data).
These choices allow us to express the macroscopic large time
average equations of the distribution of players and their
wealth explicitly in equation (\ref {Hi})

This paper is organized as follows. In Section \ref{SGTFW},
we present the multi-agent model for the dynamics of $N$ agents,
each interacting with the market (the ensemble of all agents).
This gives the Fokker-Planck equation (\ref {intro1}) for the effective single agent density $f(x,y,t)$. In Section 3, the equations are put in dimensionless form and the Gibbs measure in the frequent trading limit is introduced.  We show that the Gibbs measure expresses a Nash equilibrium, i.e. no player can improve on the cost function by choosing a different direction in $y$. In Section
  \ref{SMCL} we consider the inhomogeneous case. We introduce the GCI concept in a general setting and then, specify a simplified yet economically relevant setting where the GCI concept leads to explicit calculations. This leads to an explicit closure of the moments of the kinetic equation (\ref {intro1}). The final macroscopic model is summarized in Section \ref{Sec:macro}. Finally, we conclude by drawing some perspectives in section~\ref{SCONC}.

\section {Game theoretical framework}\label {SGTFW}

We consider a set of $N$ market agents. Each agent, labeled $j$, is endowed with two variables: its wealth $Y_j \in {\mathbb R}_+$ and a variable $X_j \in {\mathcal X}$, where  ${\mathcal X}$ is an interval of ${\mathbb R}$. The variable $X_j$ characterizes the agent's economic configuration, i.e. the category of agents it usually interacts with.
 We ignore the possibility of debts so that we take $Y_j \geq 0$.
We use notations $\vec{X}(t) =(X_1, \ldots, X_N)$, $\vec{Y}(t) =(Y_1, \ldots, Y_N)$ to describes the ensemble of all agents. To single out the market environment for the $j$-th agent, we denote
 $\hat X_j = (X_1, \ldots, X_{j-1},$ $ X_{j+1}, \ldots, X_N)$
 and $\hat Y_j = (Y_1, \ldots, Y_{j-1},$ $ Y_{j+1}, \ldots, Y_N)$ for the ensemble of all agents other than his/her self
 (note that in game theory, $\hat Y_j$ is often denoted $Y_{-j}$).
 We also write ${\vec X} = (X_j, \hat X_j)$
 and ${\vec Y} = (Y_j, \hat Y_j)$
 to represent the agent $j$ in the market environment $(X_j, \hat X_j, Y_j, \hat Y_j)$. We denote the
 cost function for the $j$-th agent in this  market environment as
 $\Phi^N(X_j, \hat X_j, Y_j, \hat Y_j, t)$  or $\Phi^N(\vec{X}, Y_j, \hat Y_j, t)$.
 The best-reply strategy is mostly used in economy.
 Each agent tries to minimize the cost function with respect to its wealth variable,
 assuming that the other agents do not change theirs.
 The agents choose the steepest descent direction of their
 cost function $Y_j \to \Phi^N(\vec{X}, Y_j, \hat Y_j)$ as their action in wealth space, i.e.,
$$
{\cal F}^N(\vec{X}, Y_j, \hat Y_j, t) = - \partial_{Y_j} \Phi^N(\vec{X}, Y_j, \hat Y_j, t)
$$
 This action is supplemented with a geometric Brownian noise which models volatility. The resulting
 dynamics of the $j$-th agent is  described below
 \begin{eqnarray}
&&\hspace{-1cm}
\dot X_j = V(X_j(t), Y_j(t)),
\label{eq:dd_wealth_X} \\
&&\hspace{-1cm}
dY_j = {\cal F}^N(\vec{X}, Y_j, \hat Y_j, t) \, dt +  \sqrt{2d} \,\,Y_j \,   dB_t^j.
\label{eq:dd_wealth_Y_2}
\end{eqnarray}
The  stochastic geometric Brownian noise is understood in the It\`o sense and
the quantity $\sqrt{2d}$ is the volatility while the notation $B_t^j$ denote independent Brownian motions.
The first equation above  describes how fast the agent evolves
in the economic configuration space as a function of its current wealth
and current economic configuration and $V(x,y)$ is a measure of the speed of this motion. We assume  that the function $V$ decays to zero at far field if the domain is unbounded, and that $V=0$ holds  on the boundary $\partial {\mathcal X}$ if the domain is bounded, i.e. 
\begin{equation}\label{eq:v-bc}
V \to 0 \quad \mbox{ as } \quad x \to \partial {\mathcal X}, 
\end{equation}
holds.

In this dynamics,  the agents would eventually, at large times, reach a point of minimum of their cost function. This minimum would then be written
\begin{eqnarray}
&&\hspace{-1cm}
Y_j^N (\vec{X}, \hat Y_j,t) = \mbox{arg} \min_{Y_j \in {\mathbb R}_+} \Phi^N(\vec{X}, Y_j, \hat Y_j, t), \quad \forall j \in \{ 1, \ldots, N\}.
\label{eq:discrete_Nash}
\end{eqnarray}
and corresponds to a Nash equilibrium of the agents. Therefore, the dynamics correspond to a
non-cooperative non-atomic anonymous game \cite{Aumann_Econometrica64, MasColell_JMathEcon84, Schmeidler_JStatPhys73, Shapiro_Shapley_MathOperRes78},
also known as a Mean-Field Game \cite{Cardaliaguet_NotesCollegeFrance12, Lasry_Lions_JapanJMath07},
where the equilibrium assumption is replaced by a time dynamics
describing the march towards a Nash equilibrium. A game theoretical framework for this general setting was developed by the authors in \cite{Degond_etal_preprint13_2} and applied to study conservative economies in \cite {Degond_etal_jsp}.

In this paper we consider a modified, and in some sense more realistic, model where the cost functional $\Phi $ does not depend on the individual values ${\hat Y}_j$ of the other agents, but depends instead on average quantities of the ensemble. This means that agents are not trading with each other individually, but trade with a market (i.e. the ensemble of all other agents), still trying to optimize their individual costs.
So we consider a cost functional of the form
$$
\Phi ^N=\Phi ^N({\vec X} ,Y_j,\Upsilon ),\quad
{\cal F}^N=-\partial _{Y_j}\Phi ^n
$$
with $\Upsilon $ given by the averaged properties of the ensemble of all agents (the market). (In this paper, we will take $\Upsilon $ to be the given by the first two moments, corresponding to the mean and the variance, of the wealth in the whole market. So,
$\Upsilon =(\Upsilon _1,\Upsilon _2)=(\sum _kY_k,\sum _kY_k^2)$
holds.)
In the limit $N \to \infty$, the one-particle distribution
function $f$ is then a solution of the  Fokker-Planck equation :
\begin{eqnarray}\label{eq:mfl_eq_finance}
&&\hspace{-1cm}
\partial_t f + \partial_x (V(x,y) f) + \partial_y (F_f \, f) = d \, \partial_y^2 \big( y^2 f \big),
\end{eqnarray}
where $F_f = F_f(x,y,t)$ is given by
\begin{eqnarray}
&&\hspace{-1cm}
F_f(x,y,t) = - \partial_y \Phi_{f(t)} (x,y) \ ,
\label{eq:mfl_force}
\end{eqnarray}
and $\Phi _f$ depends on the density $f$ only through $\Upsilon (f)$.
This equation is posed for $(x,y) \in {\mathcal X} \times [0,\infty[$. 
We supplement this equation with the no flux boundary condition at $y=0$: 
\begin{eqnarray}
&&\hspace{-1cm}
d\partial _y(y^{2}f)- F_f \, f|_{y=0}=0
, \quad \forall x \in {\mathcal X}, \quad \forall t \in {\mathbb R}_+.
\label{eq:mfl_bc_finance}
\end{eqnarray}
With the assumption (\ref{eq:v-bc}) on $V$, there is no need for any boundary condition on $f$ on $\partial {\mathcal X}$. These conditions imply that the number of agents is conserved in time for the kinetic system, i.e. 
$\int_{x \in {\mathcal X}} \int_{y \in [0,\infty)} f(x,y,t) \, dx \, dy =$ Constant. We also provide an initial condition $f(x,y,0) = f_0(x,y)$.

In this paper we consider a specific trading model with the market
and take a the following quadratic cost function with coefficients depending functionally
on the ensemble of agents
\begin{equation} \label {Phi_f}
\Phi _f(x,y) =\frac 1 2 a_f(y+\frac {b_f} {a_f}  )^2+c_f-\frac 1 2 \frac {b_f^2}{a_f}
=\frac 1 2 a_f y^2+b_f y+c_f\ ,
\end{equation}
$a_f$ represents the trading frequency with the market and
$y=-{b_f}/{a_f}$ represents the  optimum the agent  tries to achieve. Note that constant $c_f$ plays no role in
strategy ${\cal F}_f$ and we can set it as $ {b_f^2}/{(2a_f)}$.
The cost function (\ref{Phi_f}) resembles the structure of the cost function used in \cite {Degond_etal_jsp}, but contains now arbitrary coefficients $a_f$ and $b_f$. The trading frequency now is taken to be uniform and depends on the market environment. The coefficient $a_f$ will be given an interpretation in the example of the risk-adverse strategy below. The flexibility in the choice of $a_f$ and $b_f$ in the functional enables us to model market strategies. Specifically, in Section \ref{SMCL}, a risk averse strategy will be taken for $a_f$
$$
a_f=\frac {d\Upsilon _2}{\Upsilon  _2-\Upsilon  _1^2 }
$$
where $\Upsilon  _1 $ and $\Upsilon  _2$ are the first and second moments of the agent ensemble defined as above.
${a_f}/{d}$ represents the ratio between strategy action and the volatility
and is given by ${\Upsilon  _2}/{(\Upsilon  _2-\Upsilon  _1^2 )}$,
the reciprocal of the variation coefficient of the $\vec Y$.
In completely deterministic market, with no variation, the trading frequency of the agent would be infinite.
On the other hand, in an extremely uncertain market, with an infinite variance, trading frequency  would be given just by the uncertainty introduced by the Brownian motion, and $a_f=d$ holds.


\section {Dimensionless formulation and the frequent trading limit}
\label {SDFG}


\subsection {Dimensionless formulation}\label {ss3.1}
One of the main characterization in the evolution of wealth distribution
is spatio-temporal scale separation. The economic interaction
(the dynamic in the y-direction) is fast compared to the spatiotemporal
scale of the motion in the economic configuration space (i.e. the x variable).
In order to manage the various scales in a proper way,
we  change the variables to dimensionless ones.
Following the procedure developed in \cite{Degond_etal_preprint13_2}, we introduce the macroscopic scale. We assume that the changes in economic configuration $x$ are slow compared to the exchanges of wealth between agents.
We introduce $t_0$ and $x_0 = v_0 t_0$ the time and economic
configuration space units, with $v_0$ the typical magnitude of $V$.
We scale the wealth variable $y$, by a monetary unit $y_0$.
Defining $x_s=\frac {x}{x_0},\ y_s=\frac {y}{y_0}, t_s=\frac {t}{t_0}$
and $f_s(x_s,y_s,t_s)=x_0y_0f(x, y,t)$.
Correspondingly, we scale the mean wealth density $\Upsilon _1$  and the
velocity $V(x,t)$ by
$\Upsilon _1(x,t)=y_0\Upsilon _{1s}(x_s,t_s)$ and
$V(x,t)=\frac {x_0}{t_0}V_s(x_s,t_s)$.
We scale the trading frequency parameters $a_f$ and $b_f$
in (\ref {Phi}) by
$a_f=\frac 1 {\varepsilon t_0}a_{f_s}$ and $b_f=\frac {y_0} {\varepsilon t_0}b_{f_s}$
and the variance $d$ in the geometric Brownian motion
by $d=\frac {1}{\varepsilon t_0}d_s$, with $\varepsilon \ll 1 $ a small dimensionless
parameter.
This means, that we consider the frequency of the trading activity,
given by the parameters $d,a_f,b_f$ to be large compared to the
frequency of movement in the economic configuration space,
given by the average size $v_0$ of $V$.
This gives the dimensionless formulation of
 equation (\ref {intro1}) as (dropping the subscript $s$ for notational convenience):
\begin{eqnarray}
&&\hspace{-1cm}
\partial _t f^\varepsilon+\partial _x (f^\varepsilon V (x,y))=
\frac 1 \varepsilon Q(f^\varepsilon),
\label {TEsc} \\
&&\hspace{-1cm}
Q(f)=\partial _y \, [ \, d \, \partial _y (y^2 \, f)
+(a_f \, y + b_f) \, f].
\label{eq:collop}
\end{eqnarray}
In the dimensionless formulation the moment hierarchy (\ref {Hi})
is still given by
\begin{equation} \label {Hi1sc}
\partial _t \left( \begin {matrix} \rho^\varepsilon \cr \rho^\varepsilon \Upsilon _1^\varepsilon \cr \rho^\varepsilon \Upsilon _2^\varepsilon \cr \ldots
\end {matrix} \right) +
\partial _x\int V(x,y)\, f^\varepsilon(x,y,t) \left( \begin {matrix} 1\cr y \cr y^2\cr \ldots \end {matrix} \right) \ dy=
\frac 1 \varepsilon \left( \begin {matrix} 0\cr -(a_{f^\varepsilon} \Upsilon _1^\varepsilon+b_{f^\varepsilon})\cr
2(d-a_{f^\varepsilon})\Upsilon _2^\varepsilon-2b_{f^\varepsilon} \Upsilon _1^\varepsilon\cr \ldots \end {matrix} \right) \rho^\varepsilon (x,t)
 \ .
\end{equation}
The left-hand
side of (\ref {Hi1sc}) describes the slow dynamics of the moments of
distribution in the economy configuration variable $x$ and time $t$.
This evolution is driven by the fast,
local evolution of this distribution as a function of the individual decision variables $y$
described by the right-hand side. The parameter $\varepsilon$ at the denominator highlights that
fact that the internal decision variables evolve on a faster time scale than the external
economy configuration variables.
According to \cite {Degond_etal_jsp},
the fast evolution of the internal decision variables
drives
agents
performing a "rapid march", i.e. on a $O(\frac 1 \varepsilon )$ time scale,
towards a Nash equilibrium, defined by
the game of minimizing the functional $\Phi _f$ in (\ref {Phi}), up to a diffusion.

\subsection {The frequent trading limit and the Gibbs measure}
\label {ssG}
In the limit of frequent trading interaction (when $\varepsilon $
in the previous section is small compared to 1),
the macroscopic dynamics are given by the shape of the solution
of $Q(f)=0$. In the following we will restrict the form
of the nonlinear operator $Q$ such that the coefficients
$a_f$ and $b_f$  in (\ref {TEsc}) depend only on
the means of the first 2
 moments of the wealth variable. We define the vector valued functional $\underline{\Upsilon} (f)$ acting  from
the space of distribution functions into $\mathbb{R}^2$ via the definition
$$
\underline{\Upsilon} (f)= (\underline{\Upsilon}_1(f), \underline{\Upsilon}_2(f)), \quad \underline{\Upsilon} _k(f)=\frac{\int y^k f(y)\ dy}{\int f(y)\ dy} ,\
k=1,2 \ .
$$
So, the scaled trading operator $Q$ in (\ref {TEsc}) takes the form
$Q(f)=C[f,\underline{\Upsilon} (f)]$, with the operator $C$ given by
$$
Q(f)=C[f,\underline{\Upsilon} (f) ]=\partial _{y}[d\partial _{y}(y^2f)
+(a_{\underline{\Upsilon} (f) }y+ b_{\underline{\Upsilon} (f) })f].
$$
We note that, although $Q$ is a nonlinear operator, the
nonlinearity is restricted to the dependence of $Q$ on the
mean moments $\underline{\Upsilon}(f) $.
In other words, for a given vector $\Upsilon $ the operator
$C[f,\Upsilon ]$
is linear in $f$.
This allows for the definition of a normalized Gibbs measure $G_\Upsilon(y)$
satisfying (for a given vector $\Upsilon$) the linear problem
\begin{equation} \label {lG}
C[G_\Upsilon,\Upsilon]=
\partial _{y}[d\partial _{y}(y^2G_\Upsilon)
+(a_\Upsilon \, y+b_\Upsilon) \, G_\Upsilon]=0,\quad
\int _0^\infty G_\Upsilon(y)\ dy=1
\end{equation}
We reformulate the solution of $Q(f)=0$ as the combination of a
linear infinite dimensional problem (solving the linear PDE (\ref {lG})
for a given
vector $\Upsilon$), and a two dimensional fixed point problem. The computation of the local thermodynamic equilibrium, the
solution of $Q(f)=0,\ \int f\ dy=1$,
is then given by the solution $G_\Upsilon$ of (\ref {lG}) where the two-dimensional vector $\Upsilon$ is a solution of the fixed point problem:
\begin {equation}\label {l3.1}
\underline{\Upsilon} (G_\Upsilon) = \Upsilon \ .
\end {equation}

The shape of the probability distribution $f(x,y,t)$ in the
frequent trading limit $\varepsilon \rightarrow 0$ is then given
by $f^{\rm equ}(x,y,t)=\rho (x,t) G_\Upsilon(y)$,
with $G_\Upsilon$ satisfying (\ref {lG}) and $\Upsilon$ satisfying the fixed point problem
(\ref {l3.1}),
since multiplying $G_\Upsilon$ by a
$y$-independent density $\rho (x,t)$ does not change the mean moments
$\Upsilon$.

The form (\ref {lG}) of the trading operator $C[G_\Upsilon,\Upsilon]$ allows for the computation
of the mean  moment vector $\underline{\Upsilon} (G_\Upsilon)$ via a recursion formula
which is obtained by a simple integration by parts argument.
Integrating equation (\ref {lG}) against $y^k$ gives,
using the zero flux boundary condition at $y=0$
$$
\int _0^\infty [
(a_\Upsilon-d(k-1)) \, y^{k} + b_\Upsilon \, y^{k-1}] \, G_\Upsilon \ dy=0,\quad
\int _0^\infty G_\Upsilon (y)\ dy=1 \ ,
$$
and, in particular for the first two moments $\underline{\Upsilon} (G_\Upsilon)$ with $k=1,2$:
\begin{equation}
a_\Upsilon \underline{\Upsilon} _1(G_\Upsilon)+b_\Upsilon=0,\quad
(a_\Upsilon-d) \, \underline{\Upsilon} _2(G_\Upsilon)+b_\Upsilon \, \underline{\Upsilon}  _1(G_\Upsilon)=0 \ .
\label{RF}
\end{equation}
The fixed point equations (\ref {l3.1}) take then the form
\begin{equation} \label {l3.2}
a_\Upsilon \Upsilon _1+b_\Upsilon=0,\qquad
(a_\Upsilon-d)\Upsilon _2+b_\Upsilon \Upsilon _1=0 \ .
\end{equation}

\begin {itemize}
\item
So, the equilibrium solution is computed by first finding
all solutions to the fixed point equation (\ref {l3.2}),
i.e. (\ref {l3.2})
plays the role of a constitutive relation for the moments in
local equilibrium.
\item
For any vector $\Upsilon=(\Upsilon_1,\Upsilon_2)$ satisfying the
constitutive relations (\ref {l3.2}) there exists a local equilibrium
$f^{\rm equ}(x,y,t)$
given by
$f^{\rm equ}(x,y,t)=\rho (x,t) G_\Upsilon(y)$  with a local agent density $\rho (x,t)$
and $G_\Upsilon$ the solution of problem (\ref {lG}).
\item
The shape of the local equilibrium solution $f^{\rm equ}=\rho G_\Upsilon$
determines of course the large time average of the solution,
and in turn this shape depends on modeling the coefficients $a_\Upsilon$ and
$b_\Upsilon$.
So, modeling $a_\Upsilon$ and $b_\Upsilon$ determines the form of the macroscopic equations
given in Section \ref {SMCL}.
To obtain  macroscopic balance laws, in addition to the trivial conservation
law for the number of agents, the coefficients $a_Y,b_Y$ have to be
such that the constitutive relations (\ref {l3.2}) have multiple solutions.
\item
In \cite {Degond_etal_jsp} and \cite {During_Toscani_PhysicaA07}
the special case, when $a_\Upsilon$ and $b_\Upsilon$ depend only on the first
moment $\Upsilon_1$, has been treated.
In this case finding the Gibbs measure by solving (\ref {lG}),
(\ref {l3.1})
reduces to a linear problem and solutions can be computed explicitly
in terms of inverse Gamma distributions, recovering well known
results given c.f. in \cite{Amoroso_1925}.
\item
Unfortunately, it turns out that this makes the macroscopic equations trivial,
except in the case of a conservative economy when
the coefficients $a_\Upsilon$ and $b_\Upsilon$ satisfy $a_\Upsilon \, \Upsilon_1+b_\Upsilon=0$.
\item
In this paper, we therefore consider a more refined model, where
the coefficients $a_\Upsilon$ and $b_\Upsilon$ depend on $\Upsilon_1$ and $\Upsilon_2$, i.e.
on the mean and the variance of the wealth of the market,
which allows for the consideration of non - conservative economies
with $a_\Upsilon \, \Upsilon_1+b_\Upsilon\neq 0$.
\end {itemize}


\section {Large time averages and hydrodynamic hierarchy closures
using the Gibbs measure}\label {SMCL}
The goal of this section is to close the hierarchy (\ref {Hi1sc})
in Section \ref {SDFG} by a local equilibrium, i.e. by a probability density
function  $f$ of the form $f(x,y,t)=\rho (x,t)G_{\Upsilon(x,t)}(y)$ with the Gibbs measure
$G_\Upsilon(y)$ computed from the results in Subsection \ref {ssNCE}.
For a conservative economy, where the coefficients
$a_{\Upsilon },\ b_{\Upsilon }$ are such that $a_{\Upsilon }\U _1+b_{\Upsilon } =0$ holds $\forall f$
in equation (\ref {Hi}), this is rather straight
forward since we immediately obtain two conservation laws for
the density of agents and the mean wealth on large $O(\frac 1 \varepsilon )$
time scales. These can be closed by replacing $f(x,y,t)$ by the
local equilibrium density $\rho (x,t)G_{\Upsilon(x,t)} (y)$ in (\ref {Hi1sc}).
This has been done in the papers
\cite {During_Toscani_PhysicaA07} and, in a game theoretical framework,
in \cite {Degond_etal_jsp}.
\emph {In the case of a non - conservative economy
$a_{\Upsilon }\U _1+b_{\Upsilon } \neq 0$,
just taking the first moment of the transport equation \ref {TEsc}
with respect to $y$ does not yield a macroscopic conservation
law on large time scales, i.e. an equation which is independent
of $\varepsilon $.}
We therefore need to integrate the transport equation \ref {TEsc}
against a more sophisticated test function, called a generalized
collision invariant (GCI), proposed in \cite {Motsch}.

\subsection {The GCI concept}
\label {ssGCI}
We consider a kinetic equation of the form
\begin{equation} \label {l4.1}
\partial _t f^\varepsilon +\partial _x(V \, f^\varepsilon)=\frac 1 \varepsilon Q(f^\varepsilon)
\end{equation} with $Q(f)$ a nonlinear operator of the form
$Q(f)=C[f,\underline{\Upsilon} (f)]$. The mean moment operator $\underline{\Upsilon} (f)= (\underline{\Upsilon}_1(f), \ldots, \underline{\Upsilon}_K(f))$ is defined  as in Section \ref {intro} by $\int y^kf\ dy =\underline{\Upsilon} _k\int f\ dy$, $k=1, \ldots, K$. The operator
$f  \mapsto C[f,\Upsilon]$ is linear for a given
vector $\Upsilon \in {\mathbb R}_+^K$. So, the nonlinear dependence of $Q(f)$ on $f$ is restricted
to the nonlinear dependence of $C[f,\underline{\Upsilon} (f)]$ on $\underline{\Upsilon} (f)$.
Integrating (\ref {l4.1}) against any test function $z(x,y)$ w.r.t. $y$
gives
\begin{equation} \label {l4.2}
\int z\{ \partial _t f^\varepsilon +\partial _x(V \, f^\varepsilon)\} \ dy=\frac 1 \varepsilon \int z Q(f^\varepsilon)\ dy \ ,
\end{equation}
A macroscopic balance law results if $\int z Q(f)\ dy=0$.
One obvious choice is $z=1$, giving the conservation of the number of
agents.
In the case of a conservative economy, with
$\int yQ(f)\ dy=0,\ \forall f$, treated  in
\cite{During_Toscani_PhysicaA07} and \cite{Degond_etal_jsp},
the other choice is $z=y$, giving a set of hydrodynamic type equations
on the macroscopic level.
The basic idea of a GCI, developed in \cite {Motsch},
is to make the function $z$ dependent on the moments $\underline{\Upsilon} (f)$ of the
kinetic solution $f$, such that the right hand side in (\ref {l4.2})
vanishes.
This yields a macroscopic balance law of the form
\begin{equation} \label {l4.3}
\int \chi _{\underline{\Upsilon} (f^\varepsilon)} \, \{ \partial _t f^\varepsilon +\partial _x(V \, f^\varepsilon)\} \ dy=0 \ ,
\end{equation}
if, for any $\Upsilon \in {\mathbb R}_+^K$, we can find $z=\chi _\Upsilon$ such that
\begin{equation} \label {l4.3_bis}
\int \chi _\Upsilon \, C[f,\Upsilon]\ dy=0, \, \,  \forall f \, \mbox{ such that } \underline{\Upsilon} (f) = \Upsilon \, \, \mbox{ holds}.
\end{equation}
Using the special structure of $Q(f)=C[f,\underline{\Upsilon} (f)]$, this can be achieved
by using the $L^2$-adjoint of the operator $f \mapsto C[f,\Upsilon]$.
Let $C^{\rm adj}[g,\Upsilon]$ be defined by
$$
\int g \, C[f,\Upsilon]\ dy=\int f \, C^{\rm adj}[g,\Upsilon]\ dy \ .
$$
That $\chi _\Upsilon$ satisfies (\ref{l4.3_bis}) is equivalent to saying that
\begin{equation} \label {l4.9}
\exists (\lambda_1, \ldots, \lambda_K) \in {\mathbb R}^K \, \mbox{ such that } \, \,  C^{\rm adj}[\chi _\Upsilon, \Upsilon]=\sum_{k=1}^K \lambda_k (\Upsilon_k - y^k) \ .
\end{equation}
Then we have
\begin{eqnarray*}
\int \chi _{\underline{\Upsilon}(f)} \, Q(f)\ dy&=&
\int \chi _{\underline{\Upsilon}(f)} \, C[f,\underline{\Upsilon}(f)]\ dy\\&=&
\int f \, C^{\rm adj}[\chi _{\underline{\Upsilon}(f)}, \underline{\Upsilon}(f)]\ dy\\&=&
\, \sum_{k=1}^K \lambda_k \int f  \, (\underline{\Upsilon}_k(f) - y^k) \ dy = 0
 \ ,
\end{eqnarray*}
by the definition of $\underline{\Upsilon}_k(f)$.
So the problem of finding the macroscopic balance laws for
equation (\ref {l4.1})
reduces to finding all the GCI's i.e. all the solutions of (\ref{l4.9}). For any given vector $\Upsilon$, the set of associated GCI forms a linear manifold of dimension $M+1$, with $M \leq K$: indeed, the constants are solutions and form a linear space of dimension $1$ and the non-constant GCI's form a linear vector space of dimension $M$. We can have $M<K$ since some compatibility conditions between the $\lambda_k$ may be required. From now on, $\chi_\Upsilon$ denotes a vector of $M$ independent non-constant GCI.

If we can prove that the solution of the kinetic equation
(\ref {l4.1}) is really given up to order $O(\varepsilon )$
by the equilibrium solution, i.e. if $f^\varepsilon=\rho G_\Upsilon+\varepsilon f_1$ holds,
then
\begin{equation} \label {l4.6}
\partial _t(\rho \, G_\Upsilon) +\partial _x(V \rho \, G_\Upsilon)=
\frac 1 \varepsilon \rho C[G_{\underline{\Upsilon}(G_\Upsilon+\varepsilon f_1)},
\quad
\underline{\Upsilon}(G_\Upsilon+\varepsilon f_1)]
+O(\varepsilon )
\end{equation} holds. Letting  $\varepsilon \rightarrow 0$ gives an indefinite limit of the
form $\frac 0 0 $ on the right hand side of equation (\ref {l4.6}),
since $\Upsilon$ satisfies the constitutive equations $\underline{\Upsilon} (G_\Upsilon)=\Upsilon$,
and $C[G_\Upsilon,\Upsilon]=0$ holds.
Integrating (\ref {l4.6}) against $\chi _{\underline{\Upsilon} (\rho G_\Upsilon +\varepsilon f_1)}$
gives
$$
\int \chi _{\underline{\Upsilon} (\rho G_\Upsilon +\varepsilon f_1)}[\partial _t(\rho \, G_\Upsilon)
+\partial _x(V\rho \, G_\Upsilon)]\ dy=O(\varepsilon )\ ,
$$
and, in the limit $\varepsilon \rightarrow 0$ the closed macroscopic equations
\begin{equation} \label {l4.8}
\partial _t\rho +\partial _x(\rho \int V(x,y) \, G_\Upsilon \ dy)=0, \qquad
\int \chi _\Upsilon [\partial _t(\rho \, G_\Upsilon)
+\partial _x(V\rho \, G_\Upsilon)]\ dy=0\ ,
\end{equation}
with $\Upsilon$ satisfying the constitutive relations $\underline{\Upsilon} (G_\Upsilon)=\Upsilon$.

This leads to the following recipe for computing macroscopic balance
laws for a kinetic equation of the form (\ref {l4.1}) with
a collision operator $Q(f)$, only conserving the number of
agents, i.e. only satisfying $\int Q(f)\ dy=0,\ \forall f$,
but not conserving any additional moments.
\begin {itemize} \item
For a general vector $\Upsilon$, find the solution of (\ref {l4.9}).
Unfortunately, this will have to be done, in practice,  numerically
for nontrivial operators $C^{\rm adj}$.
\item
As pointed out earlier,  the Lagrange multipliers
$\lambda _k$, $k=1,  \ldots, K$
may not be chosen arbitrarily. Indeed, they have
to satisfy certain conditions, depending on the structure of the
operator $C^{\rm adj}$, such that the GCI equation (\ref {l4.9}) is solvable.
We also repeat that the GCI's form a linear vector space and that we denote by $\chi _\Upsilon$ a vector of independent non-constant GCI spanning the space of non-constant GCI.
\item
This gives in the limit $\varepsilon \rightarrow 0$
the macroscopic equations, which are independent
of the microscopic variable $y$ and the parameter $\varepsilon $:
\begin{equation} \label {l4.4}
\partial _t\rho +\partial _x(\int f \, V(x,y)\ dy )=0,\qquad
\int \chi _{\underline{\Upsilon} (f)} \, \{ \partial _t f+\partial _x (f \, V(x,y))\} \ dy=0,\
\end{equation}
with $\rho $
defined as $\rho (x,t)=\int f(x,y,t)\ dy$.
The system (\ref {l4.4}) still has to be closed by choosing an approximate
solution $f$ for the kinetic equation (\ref {l4.1}).
\item
The  system (\ref {l4.4}) is closed by choosing $f=f^{\rm equ}=\rho G_\Upsilon$,
with $G_\Upsilon$ being the Gibbs measure from Section \ref {ssNCE} in our case, this choice being justified by the formal limit $\varepsilon \to 0$ in (\ref{l4.1}).
\item
To compute the Gibbs measure $G_\Upsilon$ in Subsection \ref {ssNCE},
we have to solve the infinite dimensional problem
$C[G_\Upsilon,\Upsilon]=0,\ \int G_\Upsilon\ dy =1$,
for a general vector $\Upsilon$,
and then solve
the, finite dimensional, fixed point problem $\underline{\Upsilon} (G_\Upsilon)=\Upsilon$ for the
vector $\Upsilon$.
\item
The final macroscopic equations (\ref {l4.4}) will be of the form
\begin{equation} \label {l4.5}
\partial _t\rho +\partial _x(\int \rho G_\Upsilon \, V(x,y)\ dy )=0,\qquad
\int \chi _{\Upsilon} \, \{ \partial _t(\rho G_\Upsilon)+\partial _x(\rho G_\Upsilon \, V)\} \ dy=0,\
\end{equation} with $\Upsilon$ satisfying the constitutive relation $\underline{\Upsilon} (G_\Upsilon)=\Upsilon$.
\item
For the system (\ref {l4.5}) to be closed, the
fixed point equation  $\underline{\Upsilon} (G_\Upsilon)=\Upsilon$
should have a manifold structure, parametrized by as many independent parameters as independent non-constant GCI.
The free parameters in the fixed point equation $\underline{\Upsilon} (G_\Upsilon)=\Upsilon$
are essentially the other dependent variable (besides $\rho $)
in  the system (\ref {l4.5}), although it it might never be
explicitly expressed, but given implicitly by the constitutive
equations. In the example of the risk-adverse strategy below, the variables are the density and the mean wealth (meaning that the constitutive relation has only a one-parameter family of solutions, parametrized by the mean wealth) and the macroscopic system consists of the density conservation equation and a non-conservative balance equation for the mean-wealth.
\end {itemize}

\subsection {Non-conservative economies with risk averse
trading strategies}\label {ssNCE}
In the model, considered in this paper, individual agents
try to minimize the cost functional $\Phi _{\underline{\Upsilon}(f)} (y)$ with
$$
\Phi _\Upsilon (y)=\frac 1 2 a_\Upsilon y^2+b_\Upsilon y+c_\Upsilon
=\frac 1 2 a_\Upsilon (y+\frac {b_\Upsilon} {a_\Upsilon}  )^2+c_\Upsilon-\frac 1 2 \frac {b_\Upsilon^2}{a_\Upsilon}\ ,
$$
given market conditions represented by the density $f$.
So, $a_\Upsilon$ represents (in dimensionless variables) the frequency
of the trades with the market, i.e. the strategy of an agent
to trade or not to trade, and $y=-\frac {b_\Upsilon}{a_\Upsilon}$ represents
the (market dependent) optimum, the agent  tries to achieve.
We consider a risk averse strategy of the form
\begin{equation}
a_\Upsilon=\frac {d \, \Upsilon_2} {\Upsilon_2- \Upsilon_1^2},
\label{eq:riskadv}
\end{equation}
and refer to the end of Section \ref{SGTFW} for its interpretation. The constant in the potential does not influence the dynamics and we can take $c_\Upsilon-\frac 1 2 \frac {b_\Upsilon^2}{a_\Upsilon}=0$.
We choose the coefficient $b_\Upsilon$ such that,
\begin{equation}
b_\Upsilon = - (1 + \kappa) \, d \, \Upsilon_1,
\label{eq:bUps}
\end{equation}
with a fixed constant $\kappa >0$. This choice is motivated by the consideration of the Nash equilibrium below.

Using the choice (\ref{eq:riskadv}) for $a_\Upsilon$, we compute the Gibbs measure introduced in Section~\ref{SDFG}~\ref{ssG}
from $C[G_\Upsilon,\Upsilon]=0,\ \int G_\Upsilon\ dy=1$, i.e. from equation (\ref {lG}). It yields the constitutive relations for the vector $\Upsilon=(\Upsilon_1,\Upsilon_2)$ from
the recursion formula (\ref {RF}) as
\begin{equation}
\label {l4.10}
\frac {d\,\Upsilon _2}{\Upsilon _2-\Upsilon _1^2 } \, \Upsilon_1+b_\Upsilon=0, \qquad
(\frac {d\,\Upsilon _2}{\Upsilon _2-\Upsilon _1^2 }-d) \, \Upsilon_2 + b_\Upsilon \, \Upsilon_{1}=
\Upsilon_1 \, (\frac {d\,\Upsilon _1\, \Upsilon_2 }{\Upsilon _2-\Upsilon _1^2 }+b_\Upsilon)=0 \ .
\end{equation}
Since the two equations involved in (\ref {l4.10}) are the same, up to
a multiplicative factor $\Upsilon_1$, the first Eq. (\ref {l4.10}) yields
the constitutive relation.
For any choice of $b_\Upsilon$ (and in particular, for the choice given by (\ref{eq:bUps})), this equation
is one equation in two unknowns $\Upsilon_1,\Upsilon_2$ and has a one parameter
family of solutions.

Now, using the first equation (\ref{l4.10}) together with (\ref{eq:bUps}), we obtain
\begin{equation}
\frac {\Upsilon_2}{\Upsilon_2-\Upsilon_1^2}=-\frac {b_\Upsilon}{d\Upsilon_1}=1+\kappa , \quad \mbox{ or equivalently }  \quad \Upsilon_2 - \Upsilon_1^2 = \frac{1}{\kappa} \Upsilon_1^2.
\label{eq:constit1}
\end{equation}
This means that, at the Nash equilibrium
when every player has optimized its cost functional, there exists a finite
amount of risk in the market, measured by the fraction $\frac{1}{\kappa}$ of the squared mean wealth $\Upsilon_1^2$. So, the choice (\ref{eq:bUps}) is equivalent to
choosing some desired global risk, i.e. a global variation coefficient
$\frac 1 \kappa $ in the equilibrium market.
The first Eq. (\ref{eq:constit1}) leads to the following relation between $\Upsilon_1$ and $\Upsilon_2$ at equilibrium:
\begin{equation} \label {l4.11}
\Upsilon_2=\frac {1+\kappa }{\kappa }\Upsilon_1^2 \ .
\end{equation}
which is the form taken by the constitutive relation (\ref {l3.2}) in the present example.

To arrive at the closed macroscopic system (\ref {l4.5}) we still have
to compute the Gibbs measure $G_\Upsilon$ and the GCI $\chi _\Upsilon$ for a
general vector $\Upsilon=(\Upsilon_1,\Upsilon_2)$, satisfying the constitutive relations
(\ref {l4.11}).
The Gibbs measure is given, according to equation
(\ref {lG}) by the solution of
\begin{equation}
\partial _{y}[d \, \partial _{y}(y^2\, G_\Upsilon)
+(\frac {d \, \Upsilon_2}{\Upsilon_2-\Upsilon_1^2} \, y-d\, \Upsilon_1\, (1+\kappa ))\, G_\Upsilon]=0,\quad
\int _0^\infty G_\Upsilon(y)\ dy=1 \ ,
\label{eq:Gibbs}
\end{equation}
with $\Upsilon$ satisfying (\ref {l4.11}). Using the constitutive relations
(\ref {l4.11}) this gives
\begin{equation} \label {l4.12}
\partial _{y}[d \, \partial _{y} \, (y^2 \, G_\Upsilon)+d \, (1+\kappa ) \, (y-\Upsilon_1) \, G_\Upsilon]=0,\quad
\int _0^\infty G_\Upsilon(y)\ dy=1 \ ,
\end{equation}
together with the zero flux boundary condition
$d\,\partial _{y} \,(y^2\,G_\Upsilon)+d\,(1+\kappa )\,(y-\Upsilon_1)\,G_\Upsilon|_{y=0}=0$, which guarantees
the conservation of the number of agents in the system.
The solution of (\ref {l4.12}) is given by
\begin{equation} \label {l4.14}
G_\Upsilon(y)=\frac {1}{c_\Upsilon} \, y^{-\kappa -3} \,e^{-\frac {(1+\kappa ) \, \Upsilon_1}{y}},\quad
c_\Upsilon=\int _0^\infty y^{-\kappa -3} \, e^{-\frac {(1+\kappa )\,\Upsilon_1}{y}}\ dy \ .
\end{equation}
$G_\Upsilon$ is therefore given by an inverse Gamma distribution, i.e.
$$
G_\Upsilon(y)=g_{\kappa +2 ,(1+\kappa )\Upsilon_1 }(y)
$$
where the inverse Gamma distribution $g_{\alpha ,\beta }$ is defined as
$g_{\alpha ,\beta }=\frac {\beta ^\alpha }{\Gamma (\alpha )}y^{-1-\alpha }e^{-\frac \beta y }$
with shape parameter $\alpha $ and scale parameter $\beta $ and
$\Gamma (\alpha )$ denoting the Euler Gamma function evaluated at $\alpha$.
It is related to the  usual Gamma function $\Gamma$ by:
$\gamma _{\alpha ,\beta }(z)=\frac {\beta ^\alpha }{\Gamma (\alpha )}z^{\alpha -1}e^{-\beta z}$
by the change of variables $z = \frac 1 y $.
This distribution has been previously found in
\cite{Bouchaud_Mezard_PhypsicaA00}. When y is large, the distribution
 becomes the Pareto power law distribution,
which has a very strong agreement with
economic data (see e.g. the review in \cite{Yakovenko_Rosser_RevModPhys09}).
$G_\Upsilon(y)=g_{\kappa +2 ,(1+\kappa )\Upsilon_1 }(y)$ represent the large time average (i.e. the Nash equilibrium) of
a game of players, where each player tries to play the market
to achieve a desired risk, given by the constitutive relation
(\ref {l4.11}),
which is a dimensionless measure of the uncertainty of the market.
We also note that, in order for the local equilibrium distribution
$G_\Upsilon$ to have a finite variance, i.e. $\int _0^\infty y^2 \, G_\Upsilon \ dy<\infty $,
the value of $\kappa $ in (\ref {l4.14}) should be positive ($\kappa >0$).

\subsection{The GCI for risk-adverse trading strategies}

Let $\Upsilon = (\Upsilon_1,\Upsilon_2)$ be given, not necessarily related by the constitutive relation (\ref{l4.11}). With the choices (\ref{eq:riskadv}), (\ref{eq:bUps}), Eq. (\ref{l4.9}) is written:
\begin{eqnarray}
&&\hspace{-1cm}
 \partial_y \big( y^2 G_{\Upsilon} \partial_y \psi \big) = \lambda_1 (y - \Upsilon_1) G_{\Upsilon} + \lambda_2 (y^2 - \Upsilon_2) G_{\Upsilon}.
\label{eq:GCI2}
\end{eqnarray}
The weak formulation of this equation is
\begin{eqnarray}
&&\hspace{-1.5cm}
\int_0^\infty y^2 G_{\Upsilon} \, \partial_y \psi \, \partial_y \sigma = - \int_0^\infty \lambda_1 (y - \Upsilon_1) \, G_{\Upsilon} \, \sigma \, dy - \int_0^\infty \lambda_2 (y^2 - \Upsilon_2) \, G_{\Upsilon} \, \sigma \, dy,
\label{eq:GCI3}
\end{eqnarray}
for all $\sigma$. We note that the formalism of the paper \cite{Degond_etal_jsp} and particularly of its Lemma 3.5 applies. It uses an appropriate functional setting, and we refer the reader to\cite{Degond_etal_jsp} for the details.
In \cite{Degond_etal_jsp}, it is shown that a solution to (\ref{eq:GCI3}) exists if and only if the following solvability condition (whose necessity is easily found by taking $\sigma = 1$) is satisfied:
$$ \int_0^\infty \lambda_1 (y - \Upsilon_1) \, G_{\Upsilon} \, dy + \int_0^\infty \lambda_2 (y^2 - \Upsilon_2) \,  G_{\Upsilon} \, dy = 0, $$
or in other words:
\begin{equation}
\lambda_1 \, (\underline{\Upsilon}_1(G_{\Upsilon}) - \Upsilon_1)  + \lambda_2 (\underline{\Upsilon}_2(G_{\Upsilon}) - \Upsilon_2) = 0.
\label{eq:constraint_beta}
\end{equation}
Now, we define
\begin{equation}
\chi_\Upsilon = \frac{y^2}{2} - \Upsilon_1 \, y,
\label{eq:chiUps}
\end{equation}
Using (\ref{eq:Gibbs}) (and not (\ref{l4.12}) because we do not suppose the constitutive relation (\ref{l4.11}) to be satisfied), we get
\begin{eqnarray}
&&\hspace{-1.4cm}
 \partial_y \big( y^2 G_{\Upsilon} \partial_y \chi_\Upsilon \big)
= \frac{\Upsilon_1}{\Upsilon_2 - \Upsilon_1^2} \, \Big\{ - \Upsilon_1 (y^2 - \Upsilon_2) + \Upsilon_2 \Big( 1 + (1+\kappa) \big( 1 - \frac{\Upsilon_1^2}{\Upsilon_2} \,  \big) \Big) \, (y-\Upsilon_1) \Big\} \, G_{\Upsilon}.
\label{eq:chi}
\end{eqnarray}
This equation is of the form (\ref{eq:GCI2}) with
\begin{eqnarray*}
&&\hspace{-1cm}
\lambda_1 = \frac{\Upsilon_1}{\Upsilon_2 - \Upsilon_1^2} \, \Upsilon_2 \Big( 1 + (1+\kappa) \big( 1 - \frac{\Upsilon_1^2}{\Upsilon_2} \,  \big) \Big)  , \qquad
\lambda_2 = - \frac{\Upsilon_1}{\Upsilon_2 - \Upsilon_1^2} \,  \Upsilon_1.
\end{eqnarray*}
With the help of (\ref{RF}) to compute $\underline{\Upsilon}_k(G_{\Upsilon})$, $k=1, \, 2$, we immediately verify that the constraint (\ref{eq:constraint_beta}) is satisfied. From  (\ref{eq:constraint_beta}), it follows that the space of non-constant GCI is of dimension $1$ and since $\chi_\Upsilon$ is a non-constant GCI, all non-constant GCi are proportional to $\chi_\Upsilon$.

\subsection{The equation for the mean wealth}

Thanks to (\ref{eq:chiUps}), the second Eq. (\ref{l4.5}) is given by:
\begin{eqnarray}
&&\hspace{-1cm}
\int_0^\infty \big( \frac{y^2}{2} - \Upsilon_1(x,t) \, y \big) \, \partial_t (\rho G_\Upsilon) \, dy
+ \int_0^\infty \big( \frac{y^2}{2} - \Upsilon_1(x,t) \, y \big) \,  \, \partial_x (V(x,y) \, \rho G_\Upsilon) \, dy = 0.
\label{eq:conserv}
\end{eqnarray}
This gives
\begin{eqnarray}
&&\hspace{-1cm}
\partial_t \int_0^\infty \big( \frac{y^2}{2} - \Upsilon_1 \, y \big) \, \rho G_\Upsilon \, dy + \partial_t \Upsilon_1 \int_0^\infty \, y\, \rho G_\Upsilon \, dy \nonumber \\
&&\hspace{1cm}
+ \, \partial_x \int_0^\infty \big( \frac{y^2}{2} - \Upsilon_1 \, y \big) \, V \, \rho G_\Upsilon \, dy  + \partial_x \Upsilon_1 \int_0^\infty \, y\, V \, \rho G_\Upsilon \, dy= 0.
\label{eq:wealth_0}
\end{eqnarray}
We also remind the mass conservation equation (the first Eq. (\ref{l4.5})).  We define:
\begin{eqnarray}
&&\hspace{-1cm}
U_k(x;\Upsilon_1) =  \Big( \int_0^\infty V(x,y) \, G_\Upsilon(y)  \, y^k \,  dy \Big)\Big|_{\Upsilon_2 = \frac{1+\kappa}{\kappa} \Upsilon_1^2} , \quad k \in {\mathbb N},
\label{eq:U}
\end{eqnarray}
and we get
\begin{eqnarray}
&&\hspace{-1cm}
\partial_t \rho + \partial_x (\rho U_0) =  0\, \, .
\label{eq:mass_cons}
\end{eqnarray}
Now, we have, thanks to (\ref{l4.11}),
\begin{eqnarray}
&&\hspace{-1cm}
\partial_t \int_0^\infty \big( \frac{y^2}{2} - \Upsilon_1 \, y \big) \, \rho G_\Upsilon \, dy + \partial_t \Upsilon_1 \int_0^\infty \, y\, \rho G_\Upsilon\, dy
= - \frac{1-\kappa}{2 \kappa} \, \Upsilon_1^2 \, \partial_x (\rho U_0)  +  \frac{1}{\kappa} \rho \Upsilon_1 \partial_t \Upsilon_1 .
\label{eq:wealth_1}
\end{eqnarray}
and
\begin{eqnarray}
&&\hspace{-1cm}
\partial_x \int_0^\infty \big( \frac{y^2}{2} - \Upsilon_1 \, y \big) \, V \, \rho G_\Upsilon \, dy  + \partial_x \Upsilon_1 \int_0^\infty \, y\, V \, \rho G_\Upsilon \, dy 
= \partial_x \big( \rho \frac{U_2}{2} \big) - \Upsilon_1 \, \partial_x (\rho U_1)  .
\label{eq:wealth_2}
\end{eqnarray}
Inserting (\ref{eq:wealth_1}), (\ref{eq:wealth_2}),  into (\ref{eq:wealth_0}), we finally get the equation for the mean wealth $\Upsilon_1$:
\begin{eqnarray}
&&\hspace{-1cm}
\rho \partial_t \Upsilon_1 +  \frac{\kappa}{2 \Upsilon_1} \partial_x ( \rho U_2 ) - \big[ \kappa \, \partial_x (\rho U_1) + \frac{1-\kappa}{2} \, \Upsilon_1 \, \partial_x (\rho U_0)  \big]  = 0.
\label{eq:wealth_4}
\end{eqnarray}

\section{The macroscopic model}
\label{Sec:macro}

To summarize, the macroscopic model is the following system for the agent density $\rho(x,t)$ and the local mean wealth $\Upsilon_1(x,t)$:
\begin{eqnarray}
&&\hspace{-1cm}
\partial_t \rho + \partial_x (\rho U_0) =  0,
\label{eq:macro_1} \\
&&\hspace{-1cm}
\rho \partial_t \Upsilon_1 +  \frac{\kappa}{2 \Upsilon_1} \partial_x ( \rho U_2 ) - \big[ \kappa \, \partial_x (\rho U_1) + \frac{1-\kappa}{2} \, \Upsilon_1 \, \partial_x (\rho U_0)  \big]  = 0.
\label{eq:macro_2}
\end{eqnarray}
with
\begin{eqnarray}
&&\hspace{-1cm}
U_k = U_k(x;\Upsilon_1) =  \Big( \int_0^\infty V(x,y) \, G_\Upsilon(y)  \, y^k \,  dy \Big)\Big|_{\Upsilon_2 = \frac{1+\kappa}{\kappa} \Upsilon_1^2} , \quad k =0, \, 1, \, 2.
\label{eq:macro_U}
\end{eqnarray}
It could be further simplified by assuming specific values of $V(x,y)$. We leave this to future work.

\section{Conclusions} \label {SCONC}
We have derived a model for the large time averages
of a set of agents, interacting with each other through a
market,  and moving around in an abstract configuration space.
Each player interacts with the market ("trades") with a frequency
which is inverse proportional to the uncertainty of the market,
and tries to achieve an acceptable risk (given by a constant $\kappa $
which has to be matched to actual market data).
The model does not rely on the assumption of conservation
of the total wealth in the system, but instead uses the concept
of generalized collision invariants to derive macroscopic equations
for the large time averages.
In this sense, this paper is a generalization, as well as an alternative,
to previously considered models in
\cite {Bouchaud_Mezard_PhypsicaA00},
\cite {Degond_etal_preprint13_2},
\cite {During_Toscani_PhysicaA07},
where only binary trading interactions between individual agents
have been considered under
the assumption of conservation of the total wealth in the system.
The final macroscopic model consists of a conservation law
for the number of agents in the system and a balance law for
the mean and the variance of the total wealth, supplemented
by a constitutive relation for mean and variance.
So, in the large time limit, agents move  in configuration
space (which is assumed to be one dimensional in this paper for the sake of notational
simplicity) according to two partial differential equations
(\ref {eq:macro_1}), (\ref{eq:macro_2}) in time and one spatial variable.

\end{document}